\documentclass[a4paper, oneside, twocolumn, notitlepage, 10pt]{extarticle_ecoc}
\usepackage{ecoc}

\usepackage[T1]{fontenc}
\usepackage[utf8]{inputenc}
\usepackage[english]{babel}
\usepackage{mathtools}
\usepackage{amsfonts}
\usepackage{amsmath}
\usepackage{amssymb}
\usepackage{caption}
\usepackage{glossaries}
\usepackage{graphicx}
\usepackage{microtype}
\usepackage[normalem]{ulem}
\usepackage{siunitx}
\usepackage{xcolor}
\usepackage{setspace}

\usepackage{keyfloat}
\usepackage{tikz}
\usetikzlibrary{
	calc,
	positioning,
}

\usepackage{pgfplots}
\pgfplotsset{
	compat = newest,
	filter discard warning = false,
	every axis/.append style = {
		/pgf/number format/set thousands separator = {},
		line width = 1.0pt,
		line join = round,
	},
	major tick style = {
		thick,
		major tick length = 0.2cm,
		black,
	},
	tick align = center,
	x label style = {yshift = 0.4em},
	y label style = {yshift = -0.3em},
	legend style = {
		at = {(0.99,0.97)},
		font = \footnotesize,
		fill=white!96!black,
		fill opacity = 0.5,
		text opacity = 1,
	},
	colorbar style = {
		at = {(1.03,0.5)},
		anchor = west,
		width = 8,
		tick align = inside,
	},
}

\usepackage[capitalise]{cleveref}
\crefname{equation}{}{}

\renewcommand{\j}{\mathrm{j}}

\DeclarePairedDelimiter{\abs}{\lvert}{\rvert}

\newcommand*\dif{\mathop{}\!\mathrm{d}}

\newcommand{\mat}[1]{\mathbf{#1}}
\renewcommand{\vec}{\mathbf}

\newcommand{\group}[1]{\mathcal{\MakeUppercase{#1}}}

\newcommand{\modeS}[1]{\vec{F}_{#1}}

\newcommand{\nSec}{N_\mathrm{s}}
\newcommand{\secLen}{L_\mathrm{s}}

\newcommand{\chOut}{\vec{a}}
\newcommand{\chTM}{\mat{H}}
\newcommand{\chTMCoeff}[2]{H_{{#1}{#2}}}
\newcommand{\secTM}[1]{\mat{H}_{#1}}
\newcommand{\dispTM}{\mat{D}}
\newcommand{\coupTM}[1]{\mat{K}_{#1}}
\newcommand{\coupCoeffMat}[1]{\mat{C}_{#1}}
\newcommand{\coupCoeff}[2]{C_{{#1}{#2}}}
\newcommand{\B}{\mat{B}_0}

\newcommand{\xtTarget}{\mathrm{XT}_\mathrm{t}}
\newcommand{\xtPerGroup}[1]{\mathrm{XT}_{#1}}
\newcommand{\xtAvg}{\overline{\mathrm{XT}}}

\newacronym{sdm}{SDM}{space-division multiplexing}
\newacronym{mmf}{MMF}{multimode fiber}
\newacronym{gi}{GI}{graded-index}
\newacronym{dsp}{DSP}{digital signal processing}
\newacronym{icr}{ICR}{intermediate coupling regime}
\newacronym{ase}{ASE}{amplified spontaneous emission}
\newacronym{edfa}{EDFA}{erbium-doped fiber amplifier}
\newacronym{lmc}{LMC}{linear mode coupling}
\newacronym[longplural={power transfer matrices}]{ptm}{PTM}{power transfer matrix}
\newacronym{igsc}{IGSC}{intra-group strong coupling}
\newacronym{mplc}{MPLC}{multi-plane light converter}
\newacronym{cw}{CW}{continuous wave}

\begin{document}
\selectlanguage{english}

\title{A Simplified Model for Linear Mode Coupling in Multimode Fibers with Experimental Assessment\vspace{-0.2cm}}

\author{
    Paolo Carniello\textsuperscript{(1)}, Filipe M. Ferreira\textsuperscript{(2)}, Fabio A. Barbosa\textsuperscript{(2)}, Ming-Jun Li\textsuperscript{(3)}, Norbert Hanik\textsuperscript{(1)}
}

\maketitle

\begin{strip}
    \begin{author_descr}

        \textsuperscript{(1)} Institute for Communications Engineering, Technical University of Munich,
        \textcolor{blue}{\uline{paolo.carniello@tum.de}}

        \textsuperscript{(2)} Optical Networks Group, Dept. Electronic \& Electrical Eng., London, UK,

        \textsuperscript{(3)} Science and Technology Division, Corning Inc., Corning, NY 14831 USA,\vspace{-0.2cm}
    \end{author_descr}
\end{strip}

\renewcommand\footnotemark{}
\renewcommand\footnoterule{}

\begin{strip}
    \begin{ecoc_abstract}
        We propose a novel surrogate model for linear coupling in graded-index multimode fibers that is computationally more efficient than existing models, and can be tuned through a single measurable parameter. The model is benchmarked numerically and experimentally on a $ 90 $-polarization-mode fiber. ©2026 The Author(s)\vspace{-0.4cm}
    \end{ecoc_abstract}
\end{strip}

\section{Introduction}\vspace{-0.1cm}
\Gls{sdm} is among the architectures that are believed to be able to sustain the continuous demand for network expansion, e.g., within the context of datacenters \cite{Mosaic, zhang20}.

In order to assess the feasibility of deploying \glspl{mmf} to support such trend, computationally efficient channel models for system-level studies are needed. This helps investigate achievable information rates \cite{shtaif22}, nonlinear penalties \cite{lasagni24, carniello25}, \gls{dsp} techniques \cite{lauinger25}, and network architectures \cite{gatto24} for \gls{mmf}-based \gls{sdm} systems. In terms of linear effects, efficient models exist for both the weak and the strong coupling regimes \cite{ho13}, which assume that either no \gls{lmc} is present among different mode groups, or that all modes are strongly coupled. These regimes are extreme cases; more generally,
\glspl{mmf} operate in the \gls{icr}, which depends on the fiber design, length, and on the environment \cite{rademacher23,mazur19,ferreira17}. Models for the \gls{icr} which rely on physical assumptions  \cite{ferreira17, shemirani09,juarez14, spenner21, buch19} or on channel statistics \cite{disciullo25} have been proposed. However, the first family of models is computationally expensive, and does not always have a simple link to measurements. The model of \cite{disciullo25} addresses both issues, but it depends on a number of physical parameters and it has not been compared against the others.

In this paper we propose a surrogate model for \gls{lmc} in \gls{gi} \glspl{mmf} valid across all coupling regimes, that can be tuned
through a single measurable parameter. The model reproduces crosstalk per group trends
at a lower computational cost than existing physics-oriented models, making it a promising candidate for fast system-level and analytical studies. We also propose a methodology to compare different models and experimental results, through which we benchmark our approach against simulations and initial experiments on a $ 90 $-polarization-mode fiber.\vspace{-0.15cm}
\vspace{-0.2cm}

\section{Channel Model}
Let $ \chOut(z, f) $ be the vector of modal amplitudes at position $ z $ and frequency $ f $. Considering only linear effects, the relation between input and output of a \gls{mmf} is $ \chOut(z, f) = \chTM(z,f) \chOut(0, f) $, where the channel transfer matrix $ \chTM(z,f) $ can be expressed through a multisectional model as $ \chTM(f) = \prod_{i = 1}^{\nSec} \chTM_i (f) $ \cite{ho13}, where $ \chTM_i $ is the transfer matrix of the $ i $-th section, and $ \nSec $ is the number of sections of length $ \secLen $. Under certain assumptions \cite{ferreira17}, the frequency-dependent matrix $ \chTM_i(f) $ can be factorized as $ \chTM_i(f) = \dispTM(f) \coupTM{i} $ where $ \dispTM(f) $ is a diagonal matrix accounting for losses, and dispersion effects like modal and chromatic dispersion.
The frequency-independent matrix $ \coupTM{i} $ accounts for \gls{lmc} and it is the subject of our study. From coupled-mode theory, $ \coupTM{i} = \exp\!\left(-\j\left(\B + \coupCoeffMat{i}\right)\secLen\right) $, where $ \B $ is a diagonal matrix whose diagonal elements are the frequency-independent part of the propagation constants, which, in \gls{gi} \glspl{mmf}, have quasi-constant group-to-group distance $ \Delta\beta $, and the elements of the coupling matrix $ \coupCoeffMat{i} $ are \cite{marcuse75, palmieri25}\vspace{-0.1cm}
\begin{equation}\label{eq:coupCoeff}
    \coupCoeff{a}{b} = 2 \pi f \iint \modeS{a}^* \cdot \Delta\varepsilon \modeS{b} \, \dif A
\end{equation}
For simplicity we dropped the index $ i $, $ \modeS{a} $ is the $ a $-th modal profile (normalized as in \cite{palmieri25}), and $ \Delta\varepsilon $ is the dielectric tensor perturbation for the $ i $-th section.
 Several physics-oriented models for \gls{lmc} in \glspl{mmf} based on \cref{eq:coupCoeff} exist \cite{ferreira17, shemirani09, juarez14, spenner21, buch19}, which differ mainly for the choice of $ \Delta\varepsilon $ for each section. Some common choices include perturbations like bends \cite{shemirani09, juarez14, spenner21}, and geometrical imperfections \cite{ferreira17, spenner21}. Others assume $ \Delta \varepsilon $ to be a random process \cite{buch19}. The inclusion of splices has been proposed \cite{juarez14, spenner21}, in which case the transfer matrix is not obtained with \cref{eq:coupCoeff}, but through a projection \cite{juarez14}.

All these models require computing the modal set $ \{\modeS{a}\} $ for each specific fiber of interest, and the coupling coefficients $ \coupCoeff{a}{b} $ for each choice of type and intensity of perturbation. This process tends to be involved and to have a large setup time, in particular for those applications (like testing \gls{dsp} techniques) for which knowing of the coupling sources is not of interest. Additionally, the relation between measurable quantities and the tuning parameters has not always been made clear.

The model we propose is
\begin{equation}\label{eq:our_model}
    \coupTM{i} = e^{\j\B + \Delta \beta \, g(\xtTarget, G) \, (\mat{P}_i-\mat{P}_i^\dagger)}
\end{equation}
where $ \Delta \beta $ can be computed analytically \cite{mafi12} or fixed (e.g., we set it to $ 10^4 $), $ \mat{P}_i $ is a matrix whose elements are independent complex Gaussian random variables $ P_{ab} \sim \mathcal{CN}(0, h_{A-B}^2) $, and $ h_{A-B} $ sets the scaling of the variance of the coupling coefficients with respect to the distance $ A-B $ between the mode groups $ \group{A} $ and $ \group{B} $ to which the modes $ a $ and $ b $ belong. E.g., for a $ 90 $-mode \gls{mmf}, $ A $ and $ B $ range from $ 1 $ to $ 9 $. In the following we consider and compare two choices: 1) $ h_{A-B} = 1 $, i.e., same variance for all coupling coefficients; 2) $ h_{A-B} = e^{-2.25\abs{A-B}} $, i.e., the variance decreases with the distance between two groups. The second choice is motivated by the physics-oriented models and by the validation results. The function $ g(\xtTarget, G) $ depends on the target level of average crosstalk per section $ \xtTarget $ for a \gls{gi} \gls{mmf} with $ G $ groups. It has an analytic closed-form expression for the case $ h_{A-B} = 1 $, while we provide it numerically in \cite{gitlab_code} for the case $ h_{A-B}  = e^{-2.25\abs{A-B}}$. Note that model \cref{eq:our_model} does not require computing the modal profiles or the computation of several overlap integrals, but it relies only on the generation of Gaussian random variables, whose variance is directly tied to the desired level of crosstalk. Similar models to \eqref{eq:our_model} have been proposed earlier \cite{disciullo25,xiao14}, but they do
not depend explicitly on $ \xtTarget $.
\vspace{-0.1cm}

\section{Numerical and Experimental Validation}
The metric we use to compare the proposed model \cref{eq:our_model} against other numerical approaches and against experimental results is the crosstalk per group \cite{ferreira17}
\begin{equation}
    \xtPerGroup{A} = \frac{\sum_{b\notin \group{A}}\sum_{a \in \group{A}} \abs{\chTMCoeff{b}{a}}^2}{\sum_{a \in \group{A}}\sum_{b \in \group{A}} \abs{\chTMCoeff{b}{a}}^2}
\end{equation}
The group-averaged crosstalk per fiber, which the tuning parameter $ \xtTarget $ of model \cref{eq:our_model} is meant to match, is defined as $ \xtAvg = \frac{\sum_\group{A} \sum_{a \in \group{A}, b\notin \group{A}} \abs{\chTMCoeff{b}{a}}^2}{\sum_\group{A} \sum_{a,b \in \group{A}} \abs{\chTMCoeff{b}{a}}^2} $ \cite{mazur19}, which we observed to be well-approximated by a weighted average of $ \xtPerGroup{A} $.

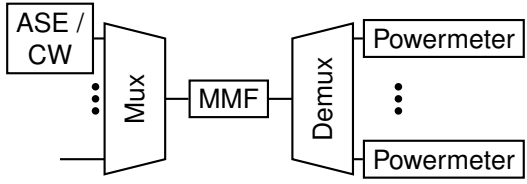
\begin{figure}[tb]
    \centering
    \def\dl{5}
\def\dyMux{2.5}
\def\dyShortMux{2}
\def\dxMux{1}
\pgfmathsetmacro{\dyyyMux}{(\dyMux-\dyShortMux)/2}
\def\xInitMux{1.5}
\def\yInitMux{-\dyMux+\dyyyMux}

\begin{tikzpicture}[
    scale = 0.8,
    thick,
    block/.style={draw, minimum width=\dl, minimum height=\dl, align=center},
    dot/.style={circle, fill, inner sep=1pt}
]

\node[block, align=center] (cw) at (0,0) {ASE /\\CW};

\node[] (cw2) at (0, -\dyShortMux) {};

\draw (0.6*\xInitMux, \yInitMux) --++ (0, \dyMux) --++ (\dxMux, -\dyyyMux) --++ (0, -\dyShortMux) coordinate[pos=0.5] (mux) -- cycle;

\node[rotate=90, anchor = center] at ($ (0.6*\xInitMux, \yInitMux) + (\dxMux/2, +\dyMux/2) $){Mux};

\draw (cw) --++ (0.6*\xInitMux, 0);
\draw (cw2) --++ (0.6*\xInitMux, 0);

\node[dot] at (\xInitMux/2, -\dyShortMux*0.4) {};
\node[dot] at (\xInitMux/2, -\dyShortMux*0.5) {};
\node[dot] at (\xInitMux/2, -\dyShortMux*0.6) {};

\node[block] (mmf) at ($(mux)+(\xInitMux*0.7, 0)$) {MMF};
\draw (mux) -- (mmf);

\draw ($(mmf)+(0.7*\xInitMux, -0.5*\dyShortMux)$) --++ (0, \dyShortMux) coordinate[pos=0.5] (demuxIn) --++ (\dxMux, \dyyyMux) --++ (0, -\dyMux) coordinate[pos=0.5] (demuxOut) -- cycle;

\node[rotate=90, anchor = center] at ($ (mmf)+(0.7*\xInitMux, -0.5*\dyShortMux) + (\dxMux/2, +\dyShortMux/2) $){Demux};

\draw (mmf) -- (demuxIn);

\node[block] (o1) at ($ (demuxOut) + (\xInitMux, 0.5*\dyShortMux) $) {Powermeter};
\draw ($(demuxOut) + (0, 0.5*\dyShortMux) $) -- (o1);

\node[block] (o2) at ($ (demuxOut) + (\xInitMux, -0.5*\dyShortMux) $) {Powermeter};
\draw ($(demuxOut) + (0, -0.5*\dyShortMux) $) -- (o2);

\node[dot] at ($ (demuxOut) + (\xInitMux/2, \dyShortMux*0.1) $) {};
\node[dot] at ($ (demuxOut) + (\xInitMux/2, \dyShortMux*0) $) {};
\node[dot] at ($ (demuxOut) + (\xInitMux/2, -\dyShortMux*0.1) $) {};

\end{tikzpicture}

    \caption{Experimental setup for the measurement of the power transfer matrix of the $ \qty{2}{\km} $ OM2 $ 90 $-mode \gls{mmf}.}
    \label{fig:experim_setup}
\end{figure}

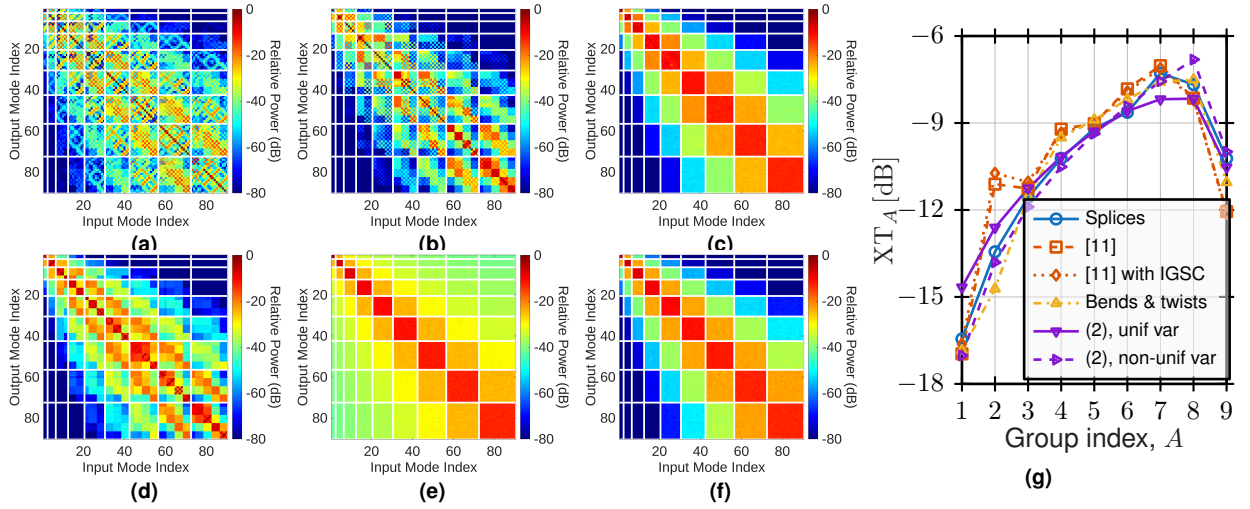
\begin{figure*}[t!]
    \centering
    \begin{minipage}{0.7\textwidth}
        \centering

        \begin{subfigure}{0.32\textwidth}
            \includegraphics[width=\linewidth]{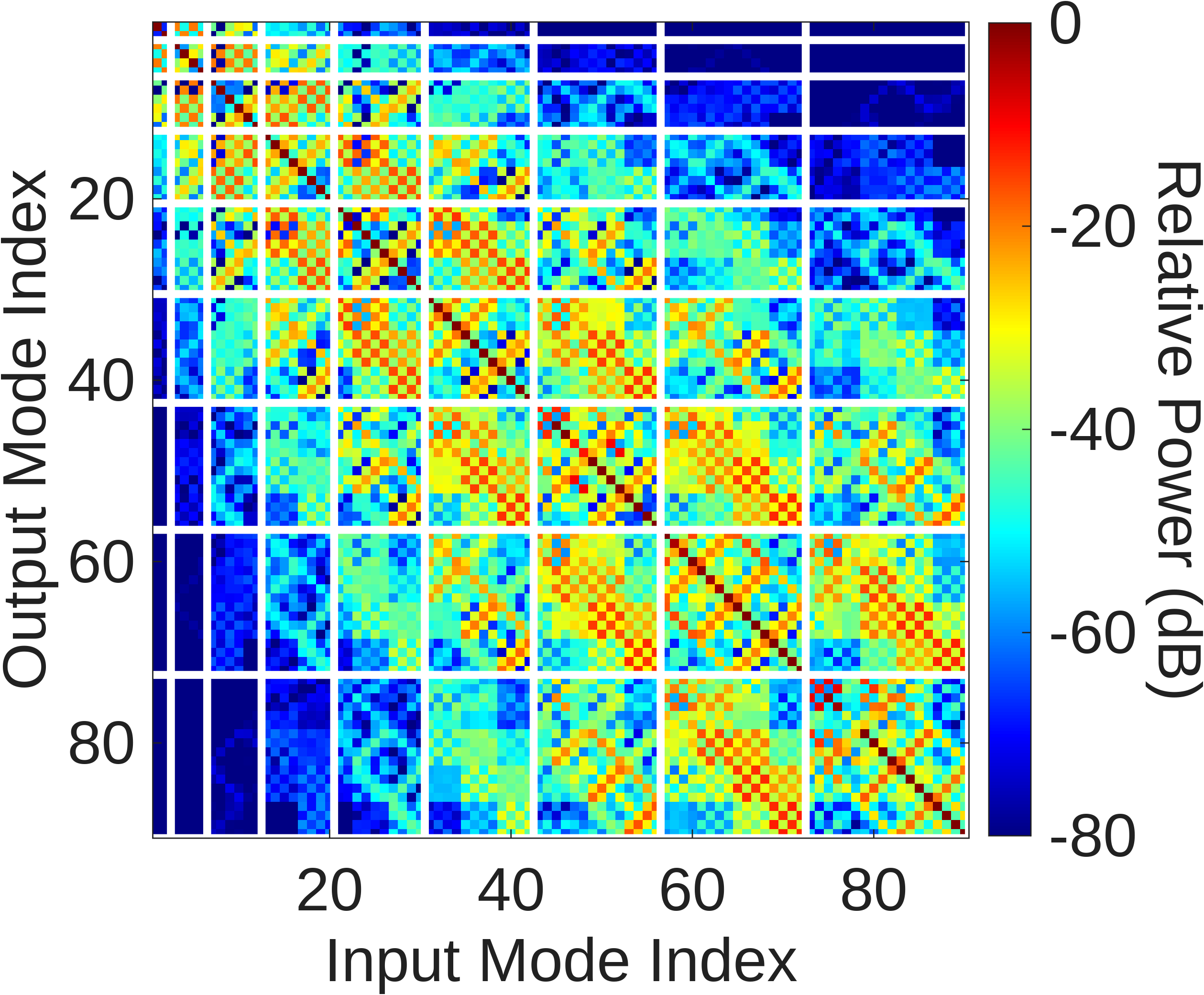}
            \caption{}
            \label{fig:splices}
        \end{subfigure}\hfill
        \begin{subfigure}{0.32\textwidth}
            \includegraphics[width=\linewidth]{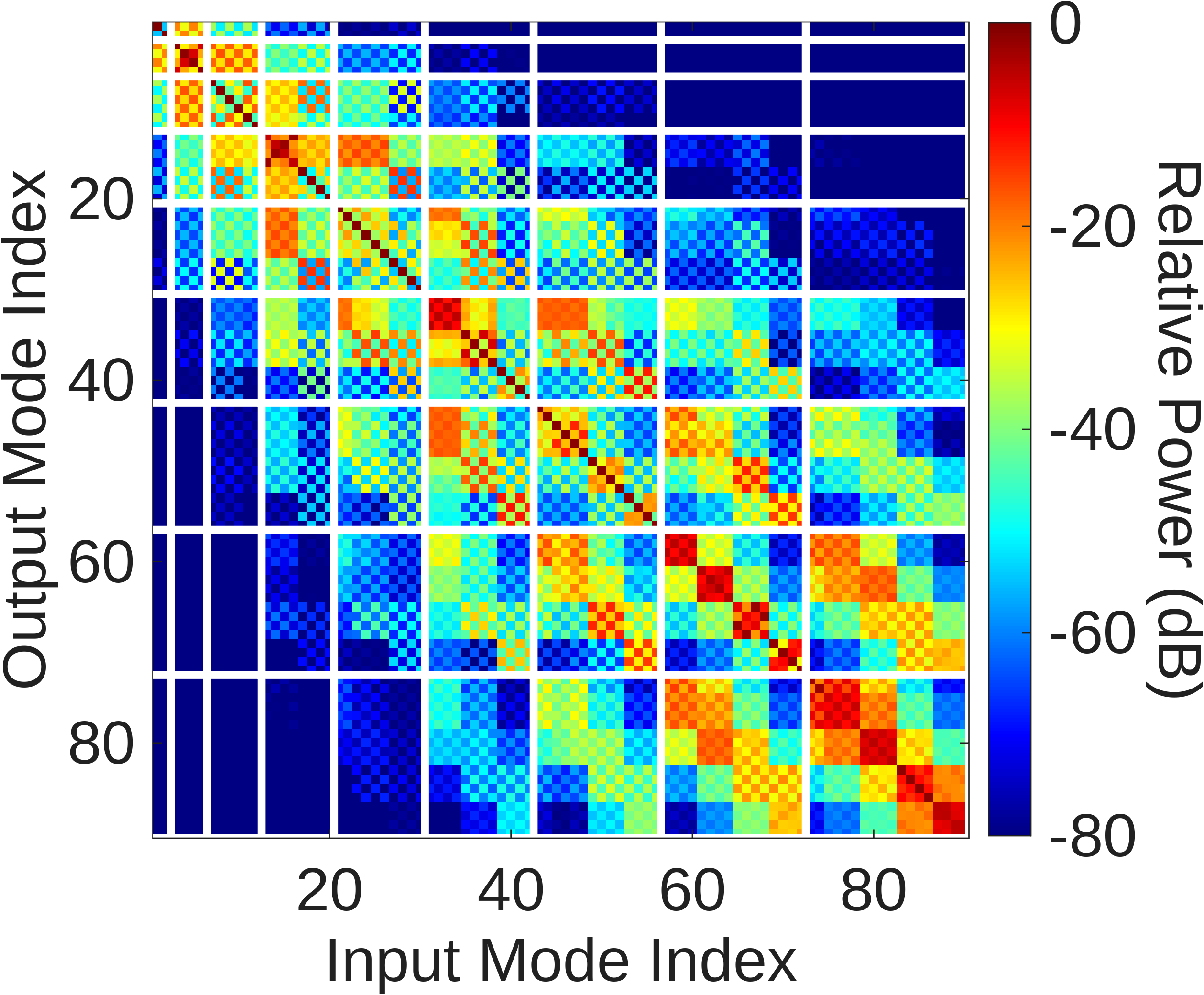}
            \caption{\label{fig:ferreira17_no_igsc}}
        \end{subfigure}\hfill
        \begin{subfigure}{0.32\textwidth}
            \includegraphics[width=\linewidth]{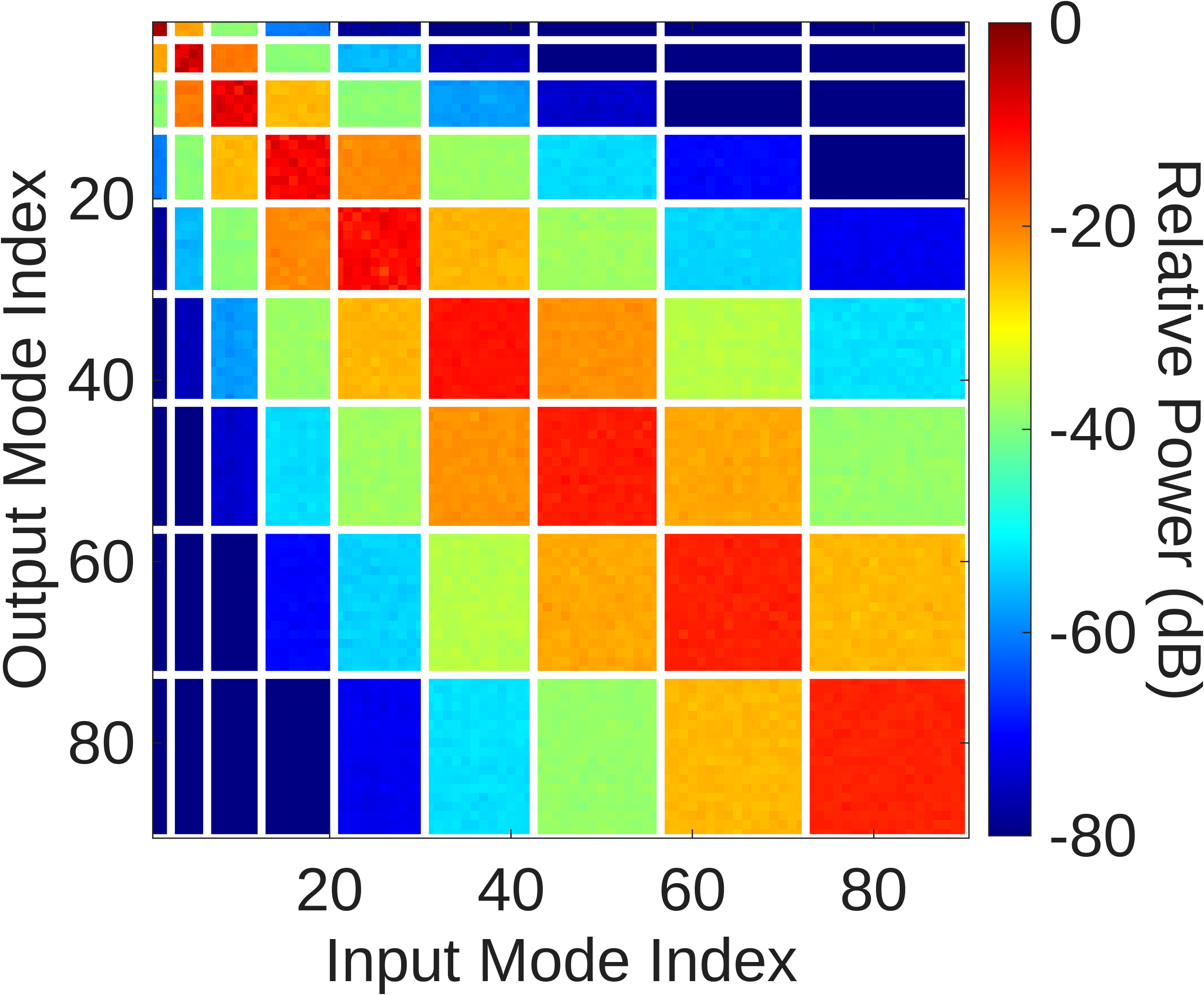}
            \caption{\label{fig:ferreira17_with_igsc}}
        \end{subfigure}

        \medskip

        \begin{subfigure}{0.32\textwidth}
            \includegraphics[width=\linewidth]{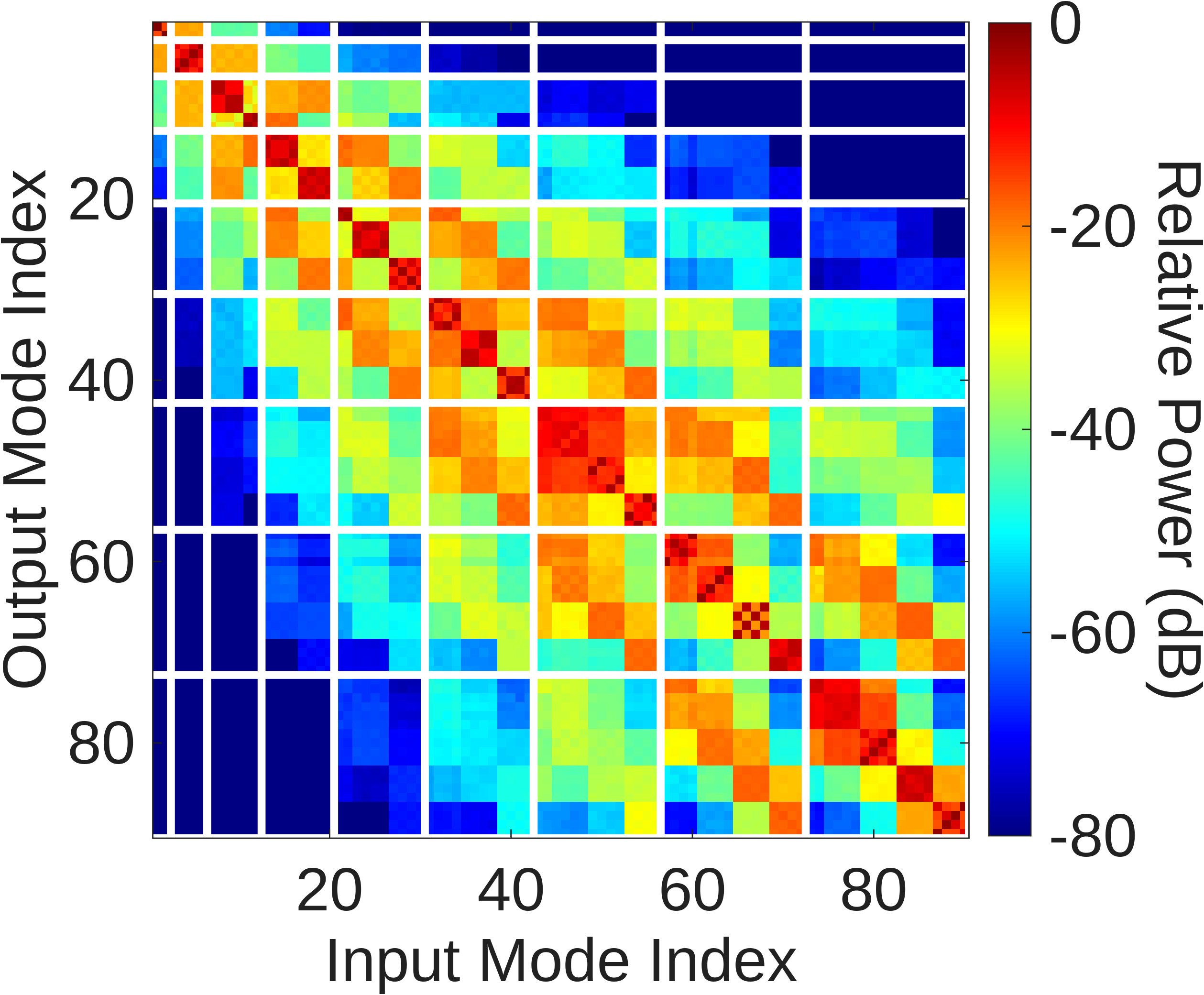}
            \caption{\label{fig:palmieri}}
        \end{subfigure}\hfill
        \begin{subfigure}{0.32\textwidth}
            \includegraphics[width=\linewidth]{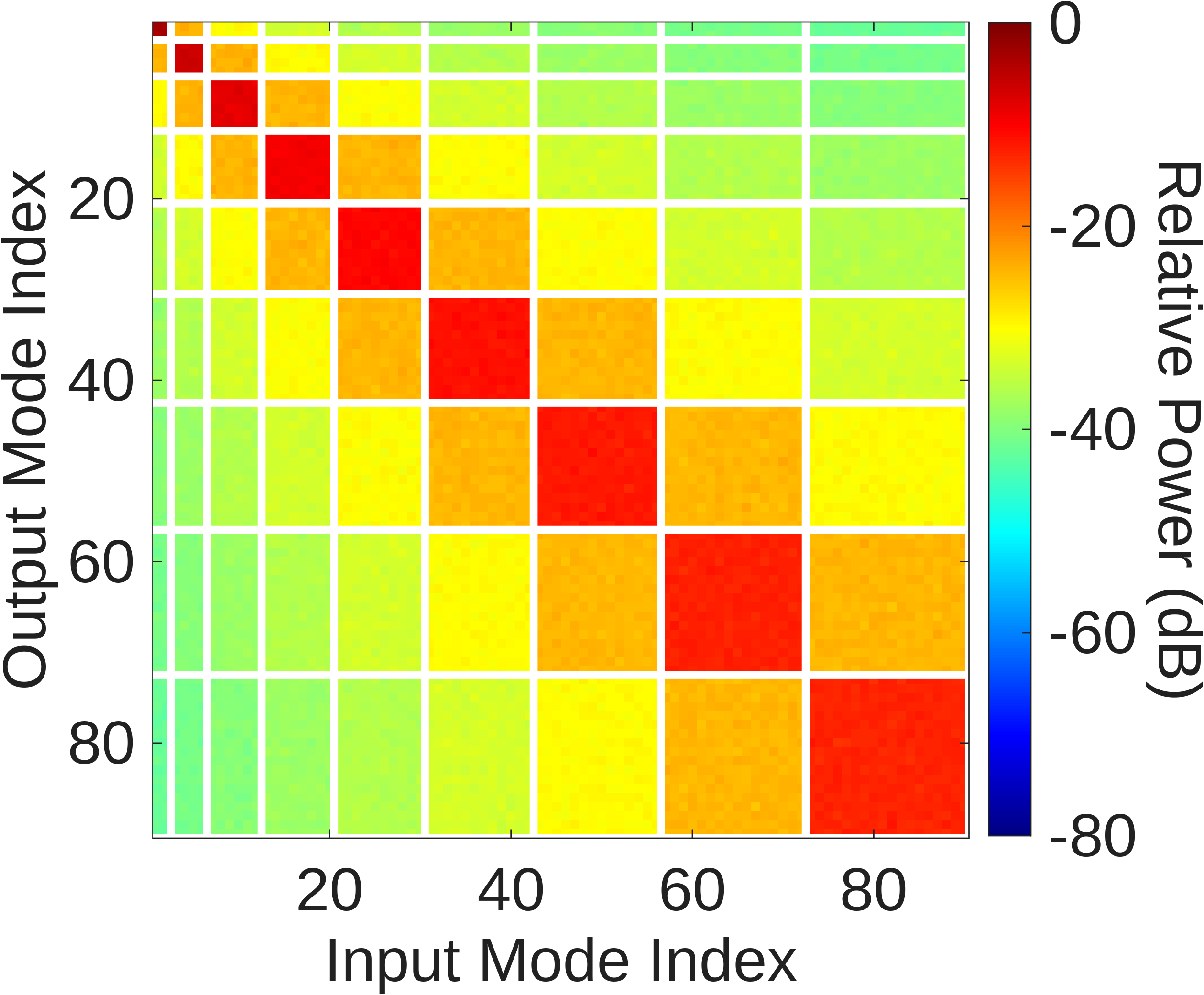}
            \caption{\label{fig:expm_unif_var}}
        \end{subfigure}\hfill
        \begin{subfigure}{0.32\textwidth}
            \includegraphics[width=\linewidth]{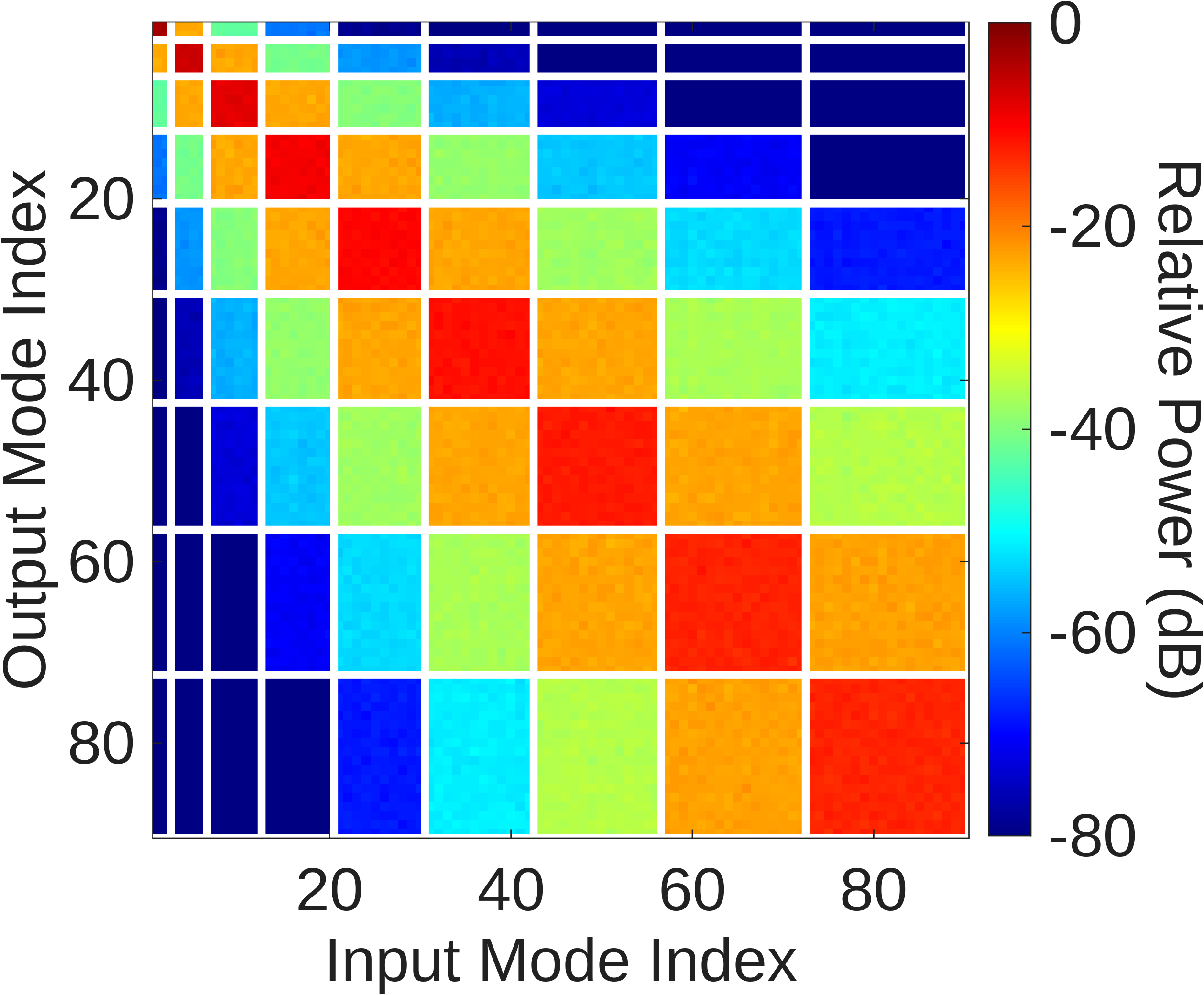}
            \caption{\label{fig:expm_v2}}
        \end{subfigure}

    \end{minipage}
    \begin{minipage}{0.29\textwidth}
        \begin{subfigure}{\linewidth}
        \centering
        \definecolor{mycolor1}{rgb}{0.06600,0.44300,0.74500}%
\definecolor{mycolor2}{rgb}{0.86600,0.32900,0.00000}%
\definecolor{mycolor3}{rgb}{0.92900,0.69400,0.12500}%
\definecolor{mycolor4}{rgb}{0.52100,0.08600,0.81900}%
\definecolor{mycolor5}{rgb}{0.12941,0.12941,0.12941}%
\begin{tikzpicture}

\begin{axis}[%
width=3.5cm,
height=4.6cm,
at={(1.888in,0.962in)},
scale only axis,
xmin=1,
xmax=9,
xtick distance=1,
xlabel style={font=\color{mycolor5}},
xlabel={Group index, $ A $},
ymin=-18,
ymax=-6,
ytick distance=3,
ylabel style={font=\color{mycolor5}},
ylabel={$ \xtPerGroup{A} \, [\unit{dB}] $},
ylabel style={yshift=-5pt},
axis background/.style={fill=white},
xmajorgrids,
ymajorgrids,
legend style={at={(0.23,0.01)}, anchor=south west, legend cell align=left, align=left, font=\scriptsize}
]
\addplot [color=mycolor1, mark=o, mark options={solid, mycolor1}]
  table[row sep=crcr]{%
1	-16.4484175016585\\
2	-13.4428319647751\\
3	-11.5664877792472\\
4	-10.2201714740885\\
5	-9.18948433994736\\
6	-8.63440537978629\\
7	-7.28389110991631\\
8	-7.67520651115129\\
9	-10.2254794251148\\
};
\addlegendentry{Splices}

\addplot [color=mycolor2, dashed, mark=square, mark options={solid, mycolor2}]
  table[row sep=crcr]{%
1	-16.9633868358971\\
2	-11.1096695227549\\
3	-11.2614866920714\\
4	-9.21131145297175\\
5	-9.02429678180851\\
6	-7.82350046952175\\
7	-7.02198512819036\\
8	-8.14419269791713\\
9	-12.0573269317958\\
};
\addlegendentry{\cite{ferreira17}}

\addplot [color=mycolor2, dotted, mark=diamond, mark options={solid, mycolor2}]
  table[row sep=crcr]{%
1	-16.6285463626046\\
2	-10.7321668055178\\
3	-11.0353987076986\\
4	-9.3438055855969\\
5	-9.00290726730952\\
6	-7.92090479824675\\
7	-7.11586563940888\\
8	-8.19779764470941\\
9	-12.1238334955855\\
};
\addlegendentry{\cite{ferreira17} with IGSC}

\addplot [color=mycolor3, dashdotted, mark=triangle, mark options={solid, mycolor3}]
  table[row sep=crcr]{%
1	-16.8454313781339\\
2	-14.718039516357\\
3	-11.5552999409072\\
4	-9.50404903641638\\
5	-8.89041396533707\\
6	-8.18248528041266\\
7	-7.58134744395944\\
8	-7.50526732351339\\
9	-11.04743527378\\
};
\addlegendentry{Bends \& twists}

\addplot [color=mycolor4, mark=triangle, mark options={solid, rotate=180, mycolor4}]
  table[row sep=crcr]{%
1	-14.6693034536422\\
2	-12.6076864292333\\
3	-11.2542069375298\\
4	-10.1641566761419\\
5	-9.28958099265678\\
6	-8.59043116065282\\
7	-8.18009423204596\\
8	-8.16116569960403\\
9	-10.527521046921\\
};
\addlegendentry{\eqref{eq:our_model}, unif var}

\addplot [color=mycolor4, dashed, mark=triangle, mark options={solid, rotate=270, mycolor4}]
  table[row sep=crcr]{%
1	-17.0189547948153\\
2	-13.8096042769998\\
3	-11.894179256061\\
4	-10.5116159656889\\
5	-9.36744019739357\\
6	-8.42260288833175\\
7	-7.56154750081188\\
8	-6.81078571437603\\
9	-9.99129803899255\\
};
\addlegendentry{\eqref{eq:our_model}, non-unif var}

\end{axis}

\end{tikzpicture}%
        \caption{\label{fig:xtPerGroup_numerical_comparison}}
    \end{subfigure}
    \end{minipage}

    \vspace{0.4cm}
    \caption{Comparison of PTMs of different models: a) splices, b) \cite{ferreira17}, c) \cite{ferreira17} with IGSC, d) bends and twists, e) Eq.~\eqref{eq:our_model} with uniform variance, f) Eq.~\eqref{eq:our_model} with non-uniform variance. g) Crosstalk per group for the six models.}
    \vspace{0.1cm}
    \label{fig:ptm_comparison}
\end{figure*}

\begin{figure*}[t!]
    \centering

    \begin{subfigure}[t]{0.24\textwidth}
        \centering
        \includegraphics[width=\linewidth]{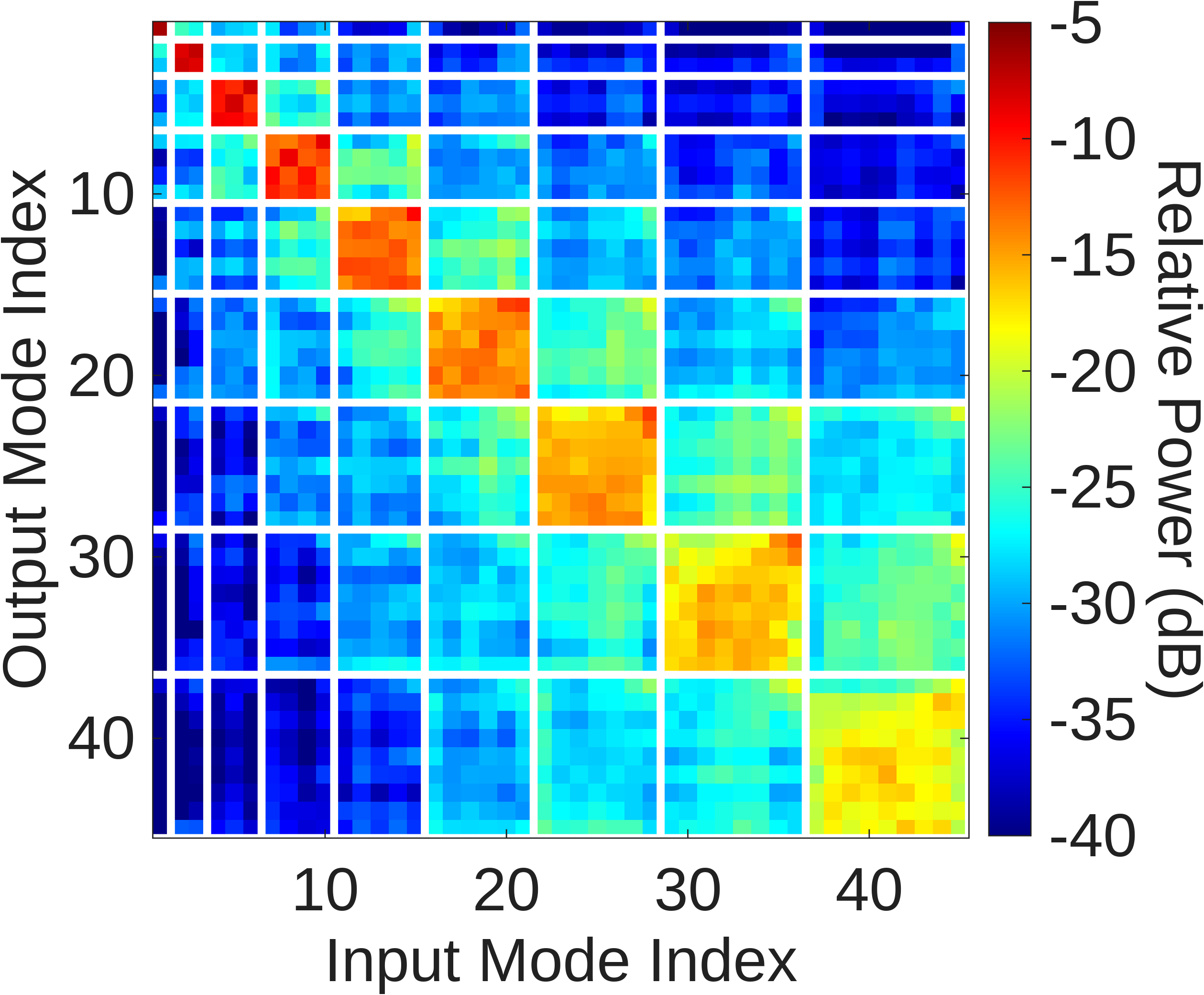}
        \caption{}
        \label{fig:xtMat_b2b}
    \end{subfigure}\hfill
    \begin{subfigure}[t]{0.24\textwidth}
        \centering
        \includegraphics[width=\linewidth]{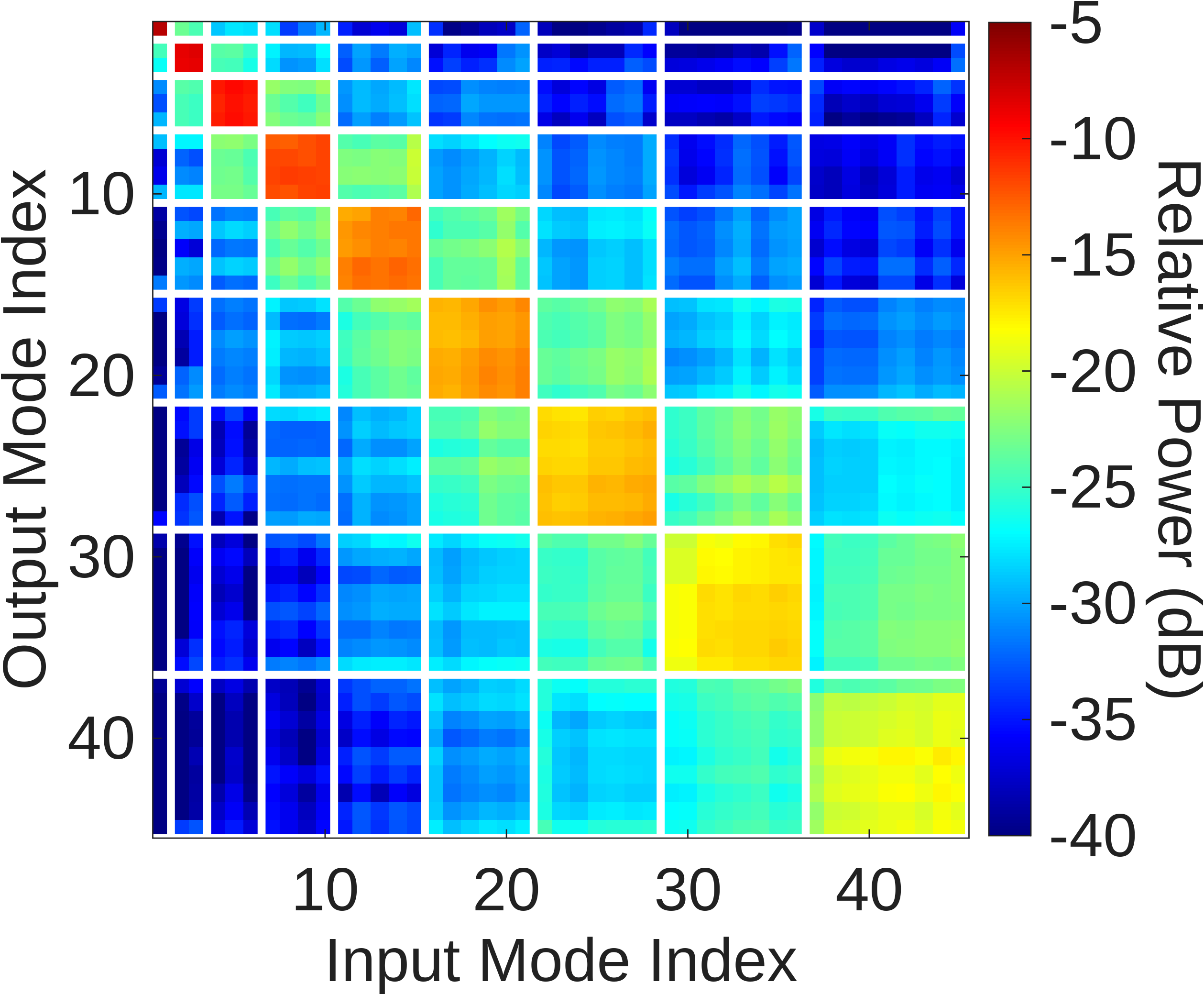}
        \caption{\label{fig:xtMat_overall}}
    \end{subfigure}\hfill
    \begin{subfigure}[t]{0.50\textwidth}
        \centering
        \definecolor{mycolor1}{rgb}{0.06600,0.44300,0.74500}%
\definecolor{mycolor3}{rgb}{0.92900,0.69400,0.12500}%
\definecolor{mycolor8}{rgb}{0.12941,0.12941,0.12941}%
\definecolor{mycolor9}{rgb}{0.86600,0.32900,0.00000}%
\begin{tikzpicture}

\begin{axis}[%
width=0.8\textwidth,
height=2.7cm,
at={(1.454in,0.741in)},
scale only axis,
xmin=1,
xmax=9,
xtick distance=1,
xlabel style={font=\color{mycolor8}},
xlabel={Group index, $ A $},
ymin=-19,
ymax=-6,
ytick={-18, -15, -12, -9, -6},
ylabel style={font=\color{mycolor8}},
ylabel={$ \xtPerGroup{A} \,\, [\unit{\dB}] $},
axis background/.style={fill=white},
xmajorgrids,
ymajorgrids,
legend style={at={(0.54, 0)}, anchor=south west, legend cell align=left, align=left, font=\scriptsize},
]

\addplot [color=mycolor3, mark=diamond, mark options={solid, mycolor3}]
  table[row sep=crcr]{%
1	-18.3093039725823\\
2	-14.4953567689856\\
3	-12.3188243176578\\
4	-10.4715522970483\\
5	-9.30129375014167\\
6	-8.64115355086635\\
7	-7.96993630459921\\
8	-8.00996911319213\\
9	-9.03131073128227\\
};
\addlegendentry{Experiment: MMF}

\addplot [color=mycolor9, dashed, mark=triangle, mark options={solid, rotate=180, mycolor9}]
  table[row sep=crcr]{%
1	-14.6693034536422\\
2	-12.6076864292333\\
3	-11.2542069375298\\
4	-10.1641566761419\\
5	-9.28958099265678\\
6	-8.59043116065282\\
7	-8.18009423204596\\
8	-8.16116569960403\\
9	-10.527521046921\\
};
\addlegendentry{\eqref{eq:our_model}, unif var}

\addplot [color=mycolor1, dashdotted, mark=triangle, mark options={solid, mycolor1}]
  table[row sep=crcr]{%
1	-17.1904832728582\\
2	-14.0059669395512\\
3	-12.0961750942441\\
4	-10.6946010983263\\
5	-9.54563191324252\\
6	-8.60122105401244\\
7	-7.75493357196913\\
8	-7.01033896417853\\
9	-10.1758847521579\\
};
\addlegendentry{\eqref{eq:our_model}, non-unif var}

\end{axis}
\end{tikzpicture}%
        \caption{\label{fig:exp_vs_model}}
    \end{subfigure}

    \vspace{0.4cm}
    \caption{Measured PTMs: a) without the fiber under test, b) with a $ \qty{2}{\km} $ OM2 \gls{mmf} with $ 90 $ polarization modes. c) Comparison in terms of crosstalk per group between the experimental results and the proposed model \eqref{eq:our_model}.\label{fig:xt_per_group_exper_comparison}}
    \label{fig:experim}
\end{figure*}
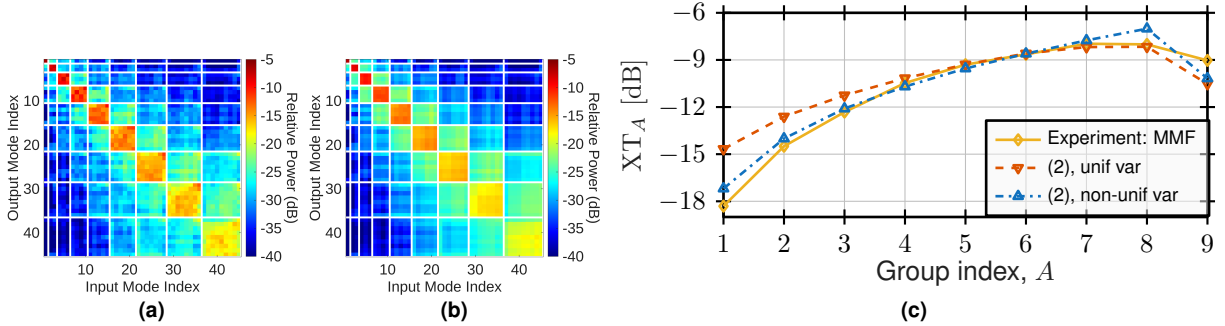

We carried out numerical validation by comparing the proposed model \cref{eq:our_model} against four reference multisectional models in which the coupling is introduced in each section by: i) a splice, i.e., $ \secTM{i} $ is computed through a projection \cite{juarez14}; ii) a superposition of bends (longitudinal and lateral strain components), and twist whose dielectric tensors $ \Delta\varepsilon $ (needed for \cref{eq:coupCoeff}) are given in \cite{palmieri25}; iii) a fluctuation in the core-cladding boundary as in \cite{ferreira17}; iv) the same transfer matrix as in the previous point multiplied by a block diagonal matrix with uniformly distributed random unitary blocks to introduce \gls{igsc}. We considered a parabolic \gls{gi} \gls{mmf} with core radius of $ \qty{25}{\micro\meter} $,
and a core-cladding contrast to guide $ 90 $ polarization modes. A solver based on \cite{fallahkhair08} has been used to compute the exact vector modes
, which were combined to obtain the LP modes \cite{palmieri25}. For each model above we considered a perturbation intensity chosen to introduce an average crosstalk per section of about $ \qty{-29}{dB} $ and enough sections ($ \sim 80$ for the concatenation of splices, $ \sim 100$ for the others) to reach $ \xtAvg \approx \qty{-9.1}{dB} $, which matches the experimental case below. The randomness among consecutive sections has been introduced by applying random rotations to the axes of consecutive section \cite{palmieri25}. The section length has been set to $ \secLen = \qty{20}{m} $ for models from ii) to iv).

\cref{fig:ptm_comparison} shows the comparison of the average \glspl{ptm} obtained with the four reference methods, and the two variations of \cref{eq:our_model}: one with $ h_{A-B}  = 1$, the other with $ h_{A-B}  = e^{-2.25\abs{A-B}}$, as mentioned above. Each matrix has been obtained averaging $ 100 $ realizations.
It can be observed in \cref{fig:expm_unif_var} that \cref{eq:our_model} with uniform variance overestimates the intensity of the blocks far away from the diagonal. Note from \cref{fig:expm_v2} that \cref{eq:our_model} with non-uniform variance reproduces more accurately the per-group intensities of the \gls{ptm}, but it cannot reproduce the specific coupling patterns between pairs of modes when \gls{igsc} is absent. When \gls{igsc} is present, which is the case in many published experimental works \cite{rademacher23, mazur19, mazur23, gatto24}, Eq.~\cref{eq:our_model} produces a similar \gls{ptm} to the reference ones, as clear from \cref{fig:expm_v2} and \cref{fig:ferreira17_with_igsc}.

\cref{fig:xtPerGroup_numerical_comparison} compares the crosstalk per group of the average \glspl{ptm} generated with the different models. They all yield comparable trends, even though differences up to a few $ \unit{dB} $ are visible. Whether or not blockwise random unitary matrices are employed to introduce \gls{igsc} has a negligible impact on $ \xtPerGroup{A} $, as seen by comparing methods iii) and iv). Note that model  \cref{eq:our_model} with non-uniform variance yields a profile that is consistent with the reference ones.

\cref{fig:experim} presents the experimental benchmarking results, for which we employed the setup displayed in \cref{fig:experim_setup}. The approach consisted in exciting with \gls{ase} noise from an \gls{edfa} each input of a $ 45 $-mode \gls{mplc} from Cailabs, whose output was connected to a spooled $ \qty{2}{\km} $ OM2 $ 90 $-mode \gls{mmf} by Corning. After the \gls{mplc} at the receiver side, power was measured simultaneously over all modes using the powermeters of a $ 96 \times 96 $ Polatis optical switch. The \gls{ptm} of the system was obtained by normalizing by the input power on each mode. We repeated the measurements with and without the fiber under test, obtaining the \gls{ptm} in \cref{fig:xtMat_b2b} for which $ \xtAvg \approx \qty{-3.9}{\dB} $, and the \gls{ptm} in \cref{fig:xtMat_overall} for which $ \xtAvg \approx \qty{-5.3}{\dB} $, respectively. The displayed \glspl{ptm}, besides being intrinsically averaged over C-band, are an average of $35$ measurements taken at $\sim 1$ minute intervals. Since the estimated average crosstalk of the fiber, $ \xtAvg \approx \qty{-9.1}{\dB} $, is lower than the crosstalk due to \gls{mplc} pair (and the splice), it was not possible to compute the average crosstalk per group $ \xtPerGroup{A} $ directly from the measured \gls{ptm}. Thus, we estimated it from the difference (in linear units) between $ \xtPerGroup{A} $ for the measurements with fiber and without fiber. We verified with numerical simulations that this method is accurate within $ \sim\qty{1}{\dB} $ in a scenario like ours. \cref{fig:xt_per_group_exper_comparison} displays the estimate of $ \xtPerGroup{A} $ for the \gls{mmf} (adjusted to compensate for mode-dependent loss), and the same $ \xtPerGroup{A} $ of \eqref{eq:our_model} as in \cref{fig:xtPerGroup_numerical_comparison}. The crosstalk per group $ \xtPerGroup{A} $ of model \cref{eq:our_model} with non-uniform variance follows the experimental one generally better than model \cref{eq:our_model} with uniform variance, with a maximum deviation from the experimental $ \xtPerGroup{A} $ below $ \qty{1.2}{\dB} $.

The measurements were repeated at $ 1 $ minute intervals for durations up to $ 12 $ hours. Even though our experiments indicate that the \gls{ptm} can have a coherence time below one minute, the crosstalk per group was observed to be stable over hours. Slight changes in the fiber position produced qualitatively similar crosstalk plots.
Slight changes in the position of the \gls{mplc} terminations produced negligible differences in crosstalk.
The measurements were repeated using a \gls{cw} source instead of the \gls{edfa} and averaging the \glspl{ptm} measured at $ 100 $ uniformly spaced frequency points in C-band. Consistent estimates of the fiber crosstalk in \cref{fig:xt_per_group_exper_comparison} were obtained.

\section{Conclusions}

We proposed a simplified model for linear coupling in graded-index multimode fibers, with a single measurable parameter to tune the desired level of coupling. The model predictions on crosstalk per group were shown to be consistent with reference models and experimental measurements from a $ 90 $-mode fiber.
This approach is suitable for further experimental evaluation of  different fibers and splicing conditions. We expect the model to simplify system-level and analytical studies of, e.g., digital signal processing and network architectures for space-division multiplexing.

\clearpage
\section{Acknowledgements}
The authors wish to thank Georg Rademacher for fruitful discussions. This work was supported by a UKRI Future Leaders Fellowship under grant no. MR/Y034260/1.

\printbibliography
\vspace{-4mm}

\end{document}